\documentclass[%aps,
reprint, prd, nofootinbib
]{revtex4-1}

\usepackage[usenames]{color}
\usepackage{subfigure,amsmath, amssymb ,amsfonts, latexsym}  
\usepackage[final]{graphicx}
\usepackage[colorlinks,hyperindex, % pagebackref,
            bookmarks=true,bookmarksopen=true]{hyperref}
    \hypersetup
    { 
        colorlinks=true,
        linkcolor=blue,
        urlcolor=blue,
        filecolor=black,
        citecolor=red,
        pdfstartview=FitV,
        pdftitle={},
        pdfauthor={Nick Evans, Keun-Young Kim, Maria Magou, Andy O'Bannon},
        pdfsubject={},
        pdfkeywords={},
        pdfpagemode=None,
        bookmarksopen=true
    }

\usepackage[notcite, color, final
                               ]{showkeys}
\definecolor{refkey}{gray}{.5}
\definecolor{labelkey}{gray}{.5}

\newcommand{\be}{\begin{equation}}
\newcommand{\ee}{\end{equation}}
\newcommand{\beqa}{\begin{subequations}\begin{eqnarray}}
\newcommand{\eeqa}{\end{eqnarray}\end{subequations}}

\newcommand{\dd}{\mathrm{d}}

\begin{document}

\newcommand\sect[1]{\emph{#1}---}

%\pagestyle{empty}

%\preprint{
%\begin{minipage}[t]{3in}
%\begin{flushright} SHEP-11-16
%\\[30pt]
%\hphantom{.}
%\end{flushright}
%\end{minipage}
%}

\title{Homes' law in Holographic Superconductor with Q-lattices}

\author{Keun-Young Kim}
\email{fortoe@gist.ac.kr}
\author{Chao Niu}
\email{chaoniu09@gmail.com}
\affiliation{  School of Physics and Chemistry, \\ Gwangju Institute of Science and Technology, Gwangju 61005, Korea
}

\begin{abstract}
\noindent
 Homes' law, $\rho_s = C \sigma_{\mathrm{DC}} T_c$, is an empirical law satisfied by various superconductors with a material independent universal constant $C$, where $\rho_{s}$ is the superfluid density at zero temperature, $T_c$ is the critical  temperature, and $\sigma_{\mathrm{DC}}$ is the electric DC conductivity in the normal state close to $T_c$. 
We study Homes' law in holographic superconductor with Q-lattices and find that Homes' law is realized for some parameter regime in insulating phase near the metal-insulator  transition boundary, where  momentum relaxation is strong. 
In computing the superfluid density, we employ two methods: one is related to the infinite DC conductivity and  the other is related to the magnetic penetration depth. With finite momentum relaxation both yield the same results, while without momentum relaxation only the latter gives the superfluid density correctly because the former has a spurious contribution from the infinite DC conductivity  due to translation invariance.  

\end{abstract}

\maketitle

%%%%%%%%%%%%%%%%%%%%%%%%%%%%%%%%%%%%%%%%%%%%%%%%%%%%%%%%%

%\acknowledgments

%\bibliography{/Users/eunyoung/Research/template/KyKimRefs}

\section{Introduction}

Holographic methods or gauge/gravity duality have provided novel and effective ways to analyse strongly correlated systems. In particular, there have been much effort and some successes in understanding 
universal properties of strongly coupled systems.  Important examples include the holographic bound of the ratio of shear viscosity to entropy density ($\eta/s$) in strongly correlated plasma, linear $T$ resistivity and Hall angle of strange metal phase~\cite{Zaanen:2015oix, Ammon:2015wua, Hartnoll:2009sz, Herzog:2009xv}.

In this paper, we study another universal property observed in high-temperature superconductors and some conventional superconductors by holographic methods. It is Home's law~\cite{Homes:2005aa,Homes:2004wv}, which connects three quantities in normal phase and condensed phase as follows:
\begin{equation} \label{H1}
\rho_{s}(T = 0) = C \sigma_{\mathrm{DC}}(T_{c}) \, T_{c} \,,
\end{equation}
where $\rho_{s}$ is the superfluid density at zero temperature, $T_c$ is the phase transition temperature, 
and $\sigma_{\mathrm{DC}}$ is the DC conductivity in the normal phase close to $T_c$. 
The point is that $C$ is a material independent universal number. $C \approx 4.4$  for ab-plane high $T_c$ superconductors and clean BCS superconductors or $C \approx 8.1$ for c-axis high $T_c$ superconductors and BCS superconductors in the dirty limit. Here, $\rho_s$, $T_c$ and $\sigma_{\mathrm{DC}}$ are defined to be dimensionless and the numerical values of $C$ are computed in~\cite{Erdmenger:2015qqa} based on the experimental data in \cite{Homes:2005aa,Homes:2004wv}.

It was argued that Homes' law might be related to `Planckian dissipation', which is the quantum limit of dissipation with  the shortest possible dissipation time
\begin{equation} \label{ppp1}
	\tau_{P} \sim \frac{\hbar}{k_B T} \,,
\end{equation}
in the normal state of high temperature superconductors~\cite{Zaanen:2004aa}.
Because the $\eta/s$ bound of strongly correlated plasma also can be explained by Planckian dissipation~\cite{Sachdev:2011cs}, Homes' law may give a good chance to find some universal physics in both condensed matter systems and quark-gluon plasma~\cite{Erdmenger:2012ik}.

Even though the holographic models of superconductor have been extensively developed~\cite{Hartnoll:2009sz, Herzog:2009xv, Horowitz:2010gk, Cai:2015cya} since the pioneering work by Hartnoll, Herzog, and Horowitz in 2008~\cite{Hartnoll:2008vx, Hartnoll:2008kx}, Homes' law in this context has not been studied much. It is partly because early holographic superconductor models are  translationally invariant with finite charge density\footnote{See \cite{Erdmenger:2012ik} for an early attempt for Homes' law in holographic superconductors without momentum relaxation.}. As a result they cannot relax momentum and yield infinite $\sigma_\mathrm{DC}$ in \eqref{H1} so $C$ is not well defined.
To have a finite $\sigma_{\mathrm{DC}}$ several methods  were proposed to incorporate momentum relaxation: spatially modulated boundary conditions for bulk fields~\cite{Horowitz:2012ky}, massive gravity models~\cite{Vegh:2013sk}, Q-lattice models~\cite{Donos:2013eha}, massless scalar models with shift symmetry~\cite{Andrade:2013gsa}, and models with a Bianchi VII$_0$ symmetry dual to helical lattices~\cite{Donos:2012js}. Based on these models, holographic superconductors incorporating  momentum relaxation have been developed~\cite{Horowitz:2013jaa, Zeng:2014uoa, Ling:2014laa, Andrade:2014xca, Kim:2015dna, Erdmenger:2015qqa, Baggioli:2015zoa,Baggioli:2015dwa, Kim:2016hzi}. 

Among the aforementioned holographic superconductors with momentum relaxation, Homes' law has been studied  only in two models \cite{Erdmenger:2015qqa,Kim:2016hzi}.  For both cases, there are parameters representing the strength of momentum relaxation, which also can be interpreted as parameters specifying material properties. Thus, Homes' law in holographic models means that $C$ is constant independent of momentum relaxation parameters.
In \cite{Erdmenger:2015qqa} a holographic superconductor model in a helical lattice was analysed and Homes' law was satisfied for some restricted parameter regime. Here the amplitude and the pitch of the helix are the momentum relaxation parameters. 
In \cite{Kim:2016hzi} a holographic superconductor model with massless scalar fields linear in spatial coordinate\footnote{The property of the normal phase and superconducting phase of this model was studied in \cite{Kim:2014bza,Kim:2015dna,Kim:2015sma,Kim:2015wba} and in \cite{Andrade:2014xca,Kim:2015dna} respectively.} are studied 
and Homes' law was not satisfied.  Here  the proportionality constant to spatial coordinate is the strength of momentum relaxation.

Therefore, it seems that Homes' law is not realized for all holographic models. Because physics behind Homes' law in  \cite{Erdmenger:2015qqa} has not been clearly understood yet, it is important to analyse other holographic models i) to see how much holographic Homes' law is robust and ii) to find the common physical mechanism for Homes' law in   different models.
For this purpose, in this paper, we study Homes' law in a holographic superconductor model with Q-lattice\footnote{The property of the normal phase of this model was studied in \cite{Donos:2013eha}. See  \cite{Ling:2014bda,Ling:2015epa} for a Mott system based on this model.}~\cite{Ling:2014laa,Andrade:2014xca}. 

We choose this model for two reasons.  First, our model can be easily compared with two previous works on Homes' law: i) the model has a similar structure to the helical lattice model~\cite{Erdmenger:2015qqa} in that it has two parameters (amplitude and wavelength of lattice) ii)  the model is also similar to the massless scalar model~\cite{Kim:2016hzi} in certain limit.  Second, it was argued in \cite{Erdmenger:2015qqa} that  Homes' law might have something to do with the metal/insulator transition in normal state and it was reported that our model also has the metal-insulator transition~\cite{Ling:2015dma}.

We find that Homes' law is realized also in our Q-lattice model for certain parameter regime, similarly to the helical lattice model in \cite{Erdmenger:2015qqa}.
However, in computing the superfluid density, there is an issue that the superfluid density is different from the charge density at zero temperature (see the end of section \ref{sec4} for more details.).
The same issue was also raised in other holographic superconductor models~\cite{Erdmenger:2015qqa, Kim:2016hzi}.
To check if the superfluid density is identified correctly, we compute superfluid density in two methods: one is related to the infinite DC conductivity and  the other is related to the magnetic penetration depth. 
Both yield the same results with finite momentum relaxation, but the only latter captures the superfluid density in the case without momentum relaxation.

This paper is organised as follows. In section \ref{sec2}, we introduce a  holographic superconductor model with Q-lattice.  
The metal-insulator transition in the normal state is also  reviewed.
In section \ref{sec3}, the superconducting transition temperature and electric DC conductivity are computed. In section \ref{sec4} the superfluid density is computed in two methods.  In section \ref{sec5} we discuss the Home's law and we  conclude in section \ref{sec6}.

\section{Holographic superconductor on a Q-lattice}\label{sec2}

In this section we briefly review a holographic superconductor model on a Q-lattice, which has been studied in detail in \cite{Ling:2014laa, Andrade:2014xca}. The action is given by 
\begin{equation}
\begin{split}
	S=\int
	\dd^4x\sqrt{-g} & \left[ R+6-\frac{1}{4}F^2 \right. \\
	& -|(\partial-iqA)\Phi|^2 -m_{\Phi}^2\Phi\Phi^{*}  \\ 
	& -|\partial
	\Psi|^2-m^2_{\Psi}|\Psi|^2 \Big], \label{eq:action}
	\end{split}
\end{equation}
where we have chosen units such that  $16\pi G = 1$ and set the AdS radius to unity. 
The first two lines are the first holographic superconductor model~\cite{Hartnoll:2008vx, Hartnoll:2008kx} with the $U(1)$ gauge field $A$, its field strength $F = \dd A$, and a complex scalar  $\Phi$.
The last line is added to introduce momentum relaxation by assuming a specific form of $\Psi$ as described below.
To be concrete, we set the mass of two scalar fields as $m_{\Psi}^2=m_{\Phi}^2=-2$.

For classical solutions we consider the following ansatz
%
%$\Psi=e^{ikx}\varphi(r)$ 
%
%A remarkable feature of such an ansatz is that while translational symmetry is broken, the gravity equations of motion are still ODEs instead of PDEs, bringing great simplifications \cite{Donos:2013eha}.
%
\begin{equation} \label{sol}
\begin{split}
&\dd s^2={1\over
	z^2}\Big[-(1-z)U(z) \dd t^2+\frac{\dd z^2}{(1-z)U(z)}  \\ 
	& \qquad \quad \quad \  + \ V_1(z) \dd x^2+V_2(z) \dd y^2\Big]\,,\\
&A=\mu(1-z)a(z) \dd t \,, \ \Phi=z\phi(z) \,, \Psi = e^{ikx}z\psi(z) \,, 
\end{split}
\end{equation}
where $U,V_1,V_2,a,\phi$ and $\psi$ are functions of only the holographic coordinate $z$. The holographic boundary is at $z=0$ and the black hole horizon is at $z=1$. The field theory temperature ($T$) is identified with the Hawking temperature $U(1)/4\pi$ with the boundary condition $U(0)=1$.
The chemical potential ($\mu$) in field theory corresponds to $A_t(0)$ with $a(0)=1$.  
The complex $\Phi$ with $m_{\Phi}^2=-2$ behaves as $\Phi = \varphi_1 z + \varphi_2 z^2 + \cdots$ near boundary. We 
choose $\varphi_1$ as a source and $\varphi_2 \equiv (\langle {\mathcal O} \rangle)$ as a condensate of the scalar operator.  For spontaneous symmetry breaking we impose the boundary condition $\varphi_1=0$. 
$\Psi$ is assumed to be the form in \eqref{sol} which breaks translation symmetry so induces momentum relaxation. It is  called Q-lattice \cite{Donos:2013eha}. With a choice $m_{\Psi}^2=-2$ the boundary value $\psi(0)=\lambda$ corresponds to the lattice amplitude and $k$ is the lattice wavenumber.

For $\Psi=\Phi=0$ the system becomes  the AdS-Reissner-Nordstr\"{o}m(AdS-RN) black hole, which allows an analytic solutions: $U=1+z+z^2-\mu^2z^3/4,~V_1=V_2=a=1,~\psi=\phi=0$.  However, for finite $\Phi$ and/or $\Psi$ we have to resort to numerical method. Our numerical solutions may be specified by four dimensionless parameters, namely $(T/\mu, \lambda/\mu, k/\mu,q)$. To be concrete, we choose $q=6$ and identify the holographic background dual to the field theory state at various $T/\mu \in (0,0.4)$ for a range of  $\lambda/\mu \in (0,90)$ and $k/\mu \in (0,20)$.

\subsection{metal-insulator transition without $\Phi$}\label{sec2a}

 We are mainly interested in properties of a holographic superconductor with $\Phi \ne 0$ in this paper. However, in this subsection, let us first consider a model with $\Psi = 0$ in \eqref{eq:action} to investigate the conductivity of our model without condensate. The result here will be used later to understand properties of a holographic superconductor.

\begin{figure}[]
	\centering
	\subfigure[\ Insulator]
	{\includegraphics[scale=0.4]{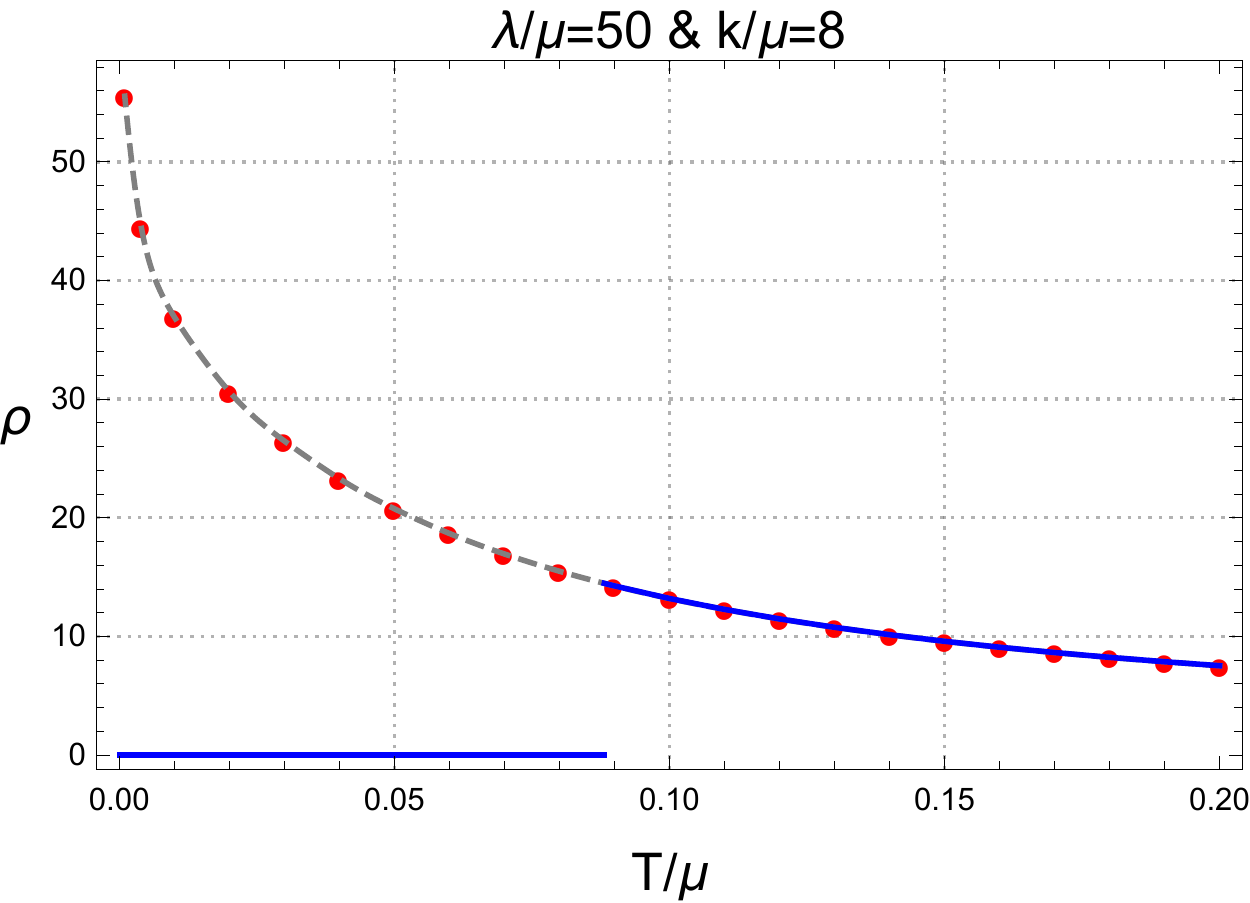}}
	\subfigure[\ Metal]
	{\includegraphics[scale=0.4]{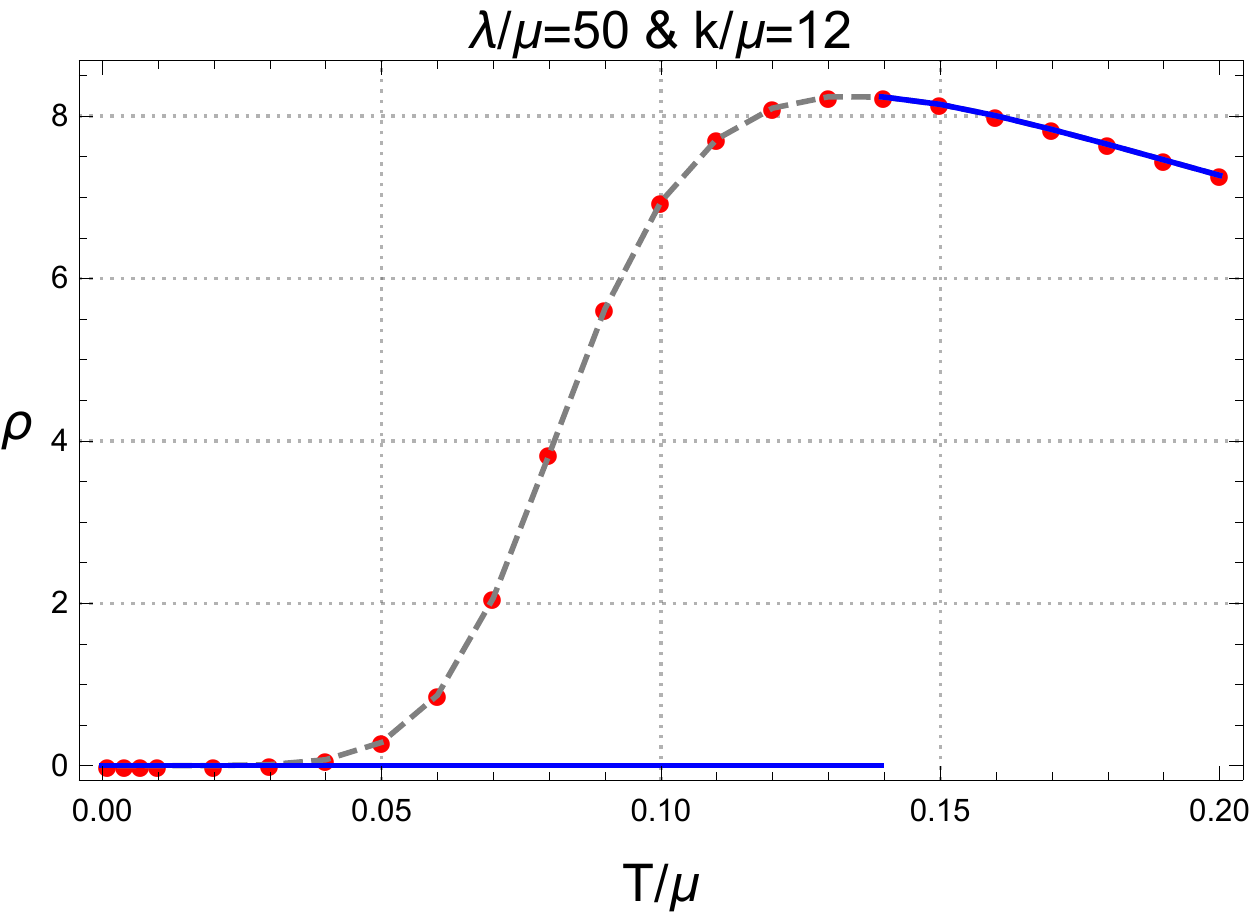}}	
	\caption{Resistivity in insulator and metal phase. Red dotted curves are the case with $\Phi=0$. 
	If $\Phi\ne0$, there is a superconducting phase transition at critical temperature $T_c$ and $\rho$ becomes zero below $T_c$. It was shown as blue lines.}\label{MIT2}
\end{figure}
\begin{figure}[]
	\centering
	\includegraphics[scale=0.4]{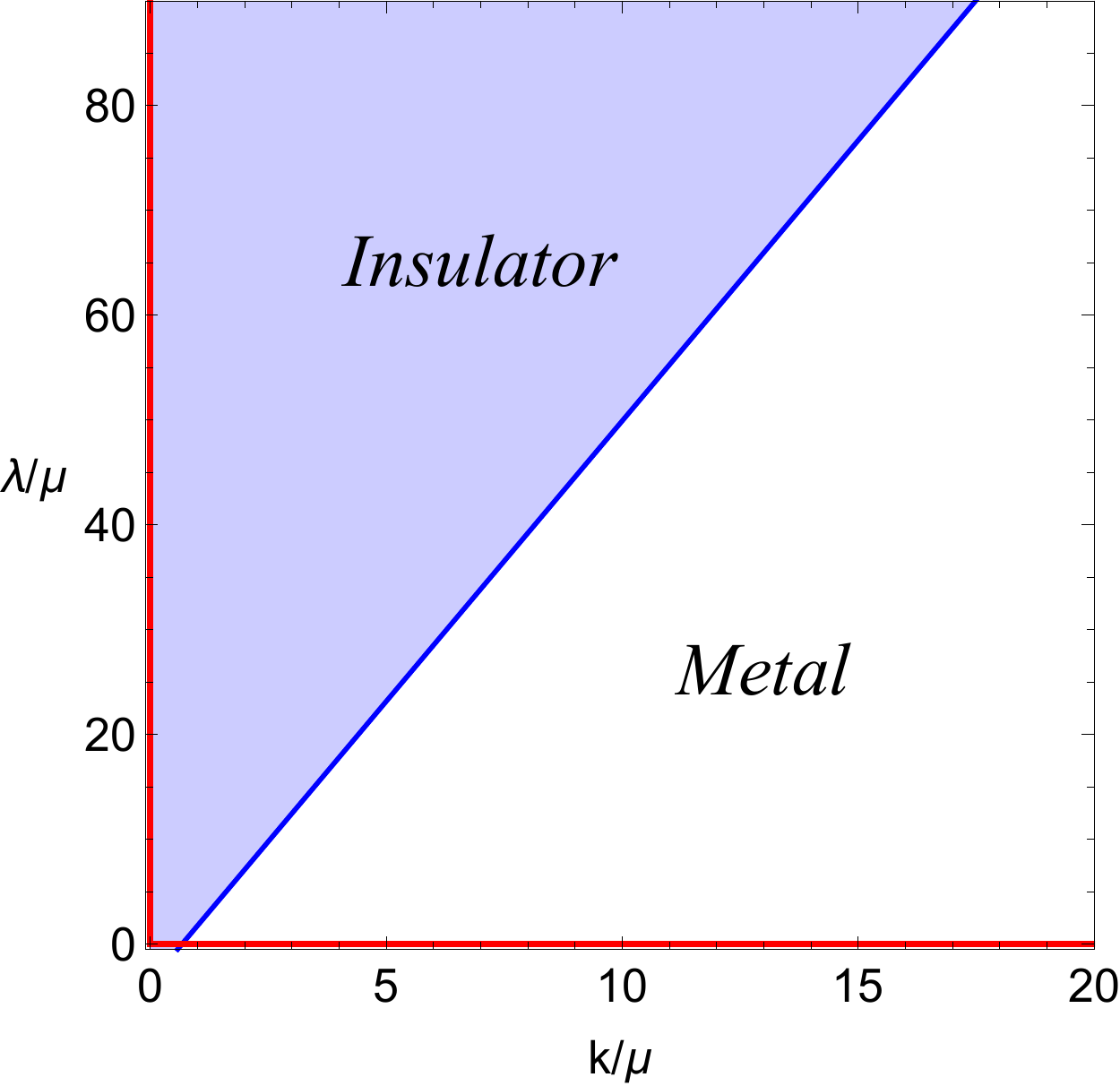}
		\caption{Metal-insulator transition with $\Phi=0$. Red lines at $k=0$ and $\lambda=0$ represent perfect metals.}\label{MIT}
\end{figure}

For our model with $\Phi=0$, it was shown that the DC conductivity, $\sigma_{\mathrm DC}$, can be computed by horizon data~\cite{Donos:2014uba}
\begin{equation} \label{dcform}
\sigma_{\mathrm{DC}}=\left. \left( \sqrt {\frac{V_2}{V_1}}  +
		\frac{{{\mu ^2}{a^2}\sqrt {{V_1}{V_2}} }}{2{k^2}{\psi ^2}}
	\right)\right|_{z = 1} \,.
\end{equation}
Plugging our numerical solutions of \eqref{sol} into \eqref{dcform} we have computed the resistivity $\rho = 1/\sigma_{\mathrm DC}$  for various values of $(T/\mu, \lambda/\mu, k/\mu)$.
For example, we show the resistivity as a function of temperature for $\lambda/\mu = 50$ in Figure \ref{MIT2}.
If $k/\mu = 8$ (a) the resistivity increases  and if $k/\mu = 12$ (b) the resistivity decreases, as temperature lowers. Therefore, the former (a) is an insulator and the latter (b) is a metal\footnote{For $\Phi \ne 0$, because of a superconducting phase transition at critical temperature $T_c$,  $\rho$ becomes zero below $T_c$ as shown by blue lines in Figure \ref{MIT2}.}. The metal insulator transition occurs at $k/\mu \approx 10.1$.  By considering several values of $\lambda/\mu$ and $k/\mu$ we obtained a phase diagram for metal-insulator transition (MIT), which is shown in Figure \ref{MIT}\footnote{{This phase diagram was first studied in \cite{Ling:2015dma} and here we extended the analysis for a much bigger range of $\lambda/\mu$ and $k/\mu$ to explore Homes' law in a big enough parameter space.}}. If $k=0$ or $\lambda=0$ (red lines) translation symmetry is recovered and the system becomes perfect metal without momentum relaxation. 

{MIT can be understood also by \eqref{dcform}.  For small $k$ (insulating phase), as temperature lowers it turns out $V_2(1)$ goes to zero, which yields $\sigma_{DC} \rightarrow 0$. Because the entropy of the system is $4\pi \sqrt{V_1(1) V_2(1)}$, the entropy vanishes in insulating phase. For large $k$ (metal phase), $\psi(1)$ goes to zero similarly to Figure \ref{fig3}, which yields a large $\sigma_{\mathrm{DC}}$. In metal phase, the entropy is finite. }

\section{Critical temperature and DC conductivity} \label{sec3}

To study Homes' law we need three quantities, critical temperature $T_c$, DC conductivity at $T_c$ ($\sigma_{\mathrm{DC}}(T_c)$) and superfluid density. 
In this section we compute the first two and in the next section we investigate superfluid density in more detail.

Using pseudo spectral method~\cite{Zhang:2016coy}, we numerically constructed classical solutions \eqref{sol} for various three dimensionless parameters $(T/\mu, \lambda/\mu, k/\mu)$ and $q=6$.  For every set of parameters $(\lambda/\mu,k/\mu)$, there is a solution with $\Phi=0$ (normal state). In addition we find another solution with $\Phi \ne 0$ (superconducting state) below the critical temperature $T_c/\mu$. In this case, the superconducting state has lower free energy than normal state so a phase transition occurs at $T_c/\mu$. 

\begin{figure}[]
	\centering
	\includegraphics[scale=0.65]{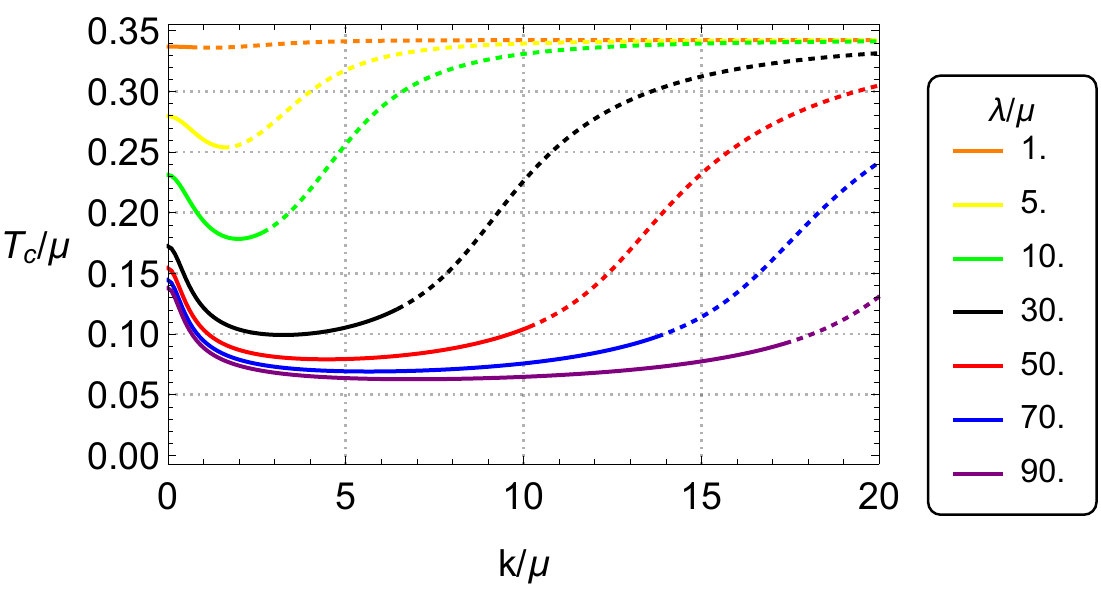}
	\caption{Critical temperature ($T_c/\mu$) vs lattice wavenumber ($k/\mu$) at fixed lattice amplitude ($\lambda/\mu=1,5,10,30,50,70,90$). The solid part and dotted part correspond to insulator and metal respectively in Figure \ref{MIT}. }\label{fig1}
\end{figure}

In Figure \ref{fig1} we illustrate how the critical temperature depends on $\lambda/\mu$ and $k/\mu$. First, for a fixed $k/\mu$, the critical temperature decreases monotonically with the increase of $\lambda/\mu$. Second, for a fixed $\lambda/\mu$, the critical temperature first decreases for small $k/\mu$, and then increases for large  $k/\mu$. As $k/\mu \rightarrow \infty$, it approaches to the critical temperature of the AdS-RN ($\lambda =0$).  A similar non-monotonic behaviour
was also observed in the massless scalar model \cite{Kim:2015dna} and the helical lattice model \cite{Erdmenger:2015qqa}. However, this behaviour was not seen in the previous analysis of Q-lattice models \cite{Ling:2014laa, Andrade:2014xca}, where the scalar field $\Phi$ has a smaller charge $q=2$ than our case ($q=6$).

\begin{figure}[]
	\centering
	\includegraphics[scale=0.6]{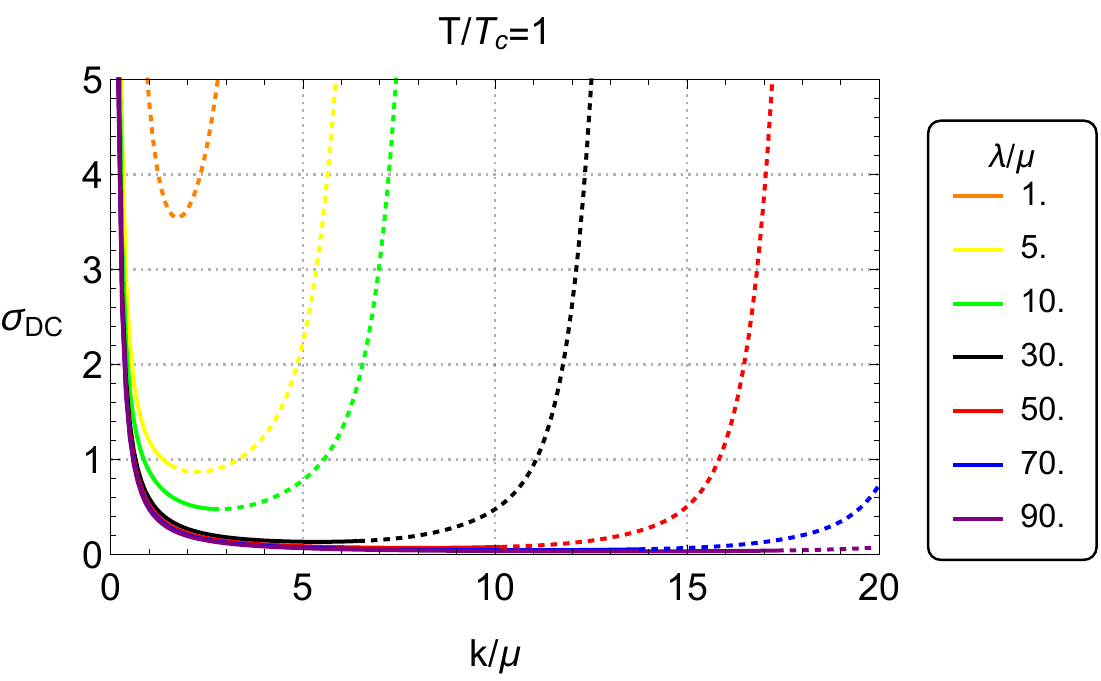}
	\caption{DC conductivity at $T_c$ ($\sigma_{\mathrm{DC}}$) vs lattice wavenumber ($k/\mu$) at fixed lattice amplitude ($\lambda/\mu=1,5,10,30,50,70,90)$. The solid part and dotted part correspond to insulator and metal respectively in Figure \ref{MIT}.}\label{fig2}
\end{figure}

Next, we compute the conductivity at $T_c$ ($\sigma_{\mathrm{\mathrm{DC}}}(T_c)$) for a range of $\lambda/\mu$ and $k/\mu$.
We use the formula \eqref{dcform} and our results are shown in Figure \ref{fig2}. When $k/\mu=0$, we have infinite $\sigma_{\mathrm{DC}}$ because the system is translationally invariant. For a fixed $\lambda/\mu$, $\sigma_{\mathrm{DC}}(T_c)$  decreases when $k/\mu$ is small and increases when $k/\mu$ is large. As $k/\mu \rightarrow \infty$, it again goes to infinity.

%%%%%%%%%%%%%%%%%%%%%%%%%%%%%%%%%%%%%%%%%
%%%%%%%%%%%%%%%%%%%%%%%%%%%%%%%%%%%%%%%%%
%%%%%%%%%%%%%%%%%%%%%%%%%%%%%%%%%%%%%%%%%
%%%%%%%%%%%%%%%%%%%%%%%%%%%%%%%%%%%%%%%%%

\begin{figure}[]
	\centering
	\includegraphics[scale=0.45]{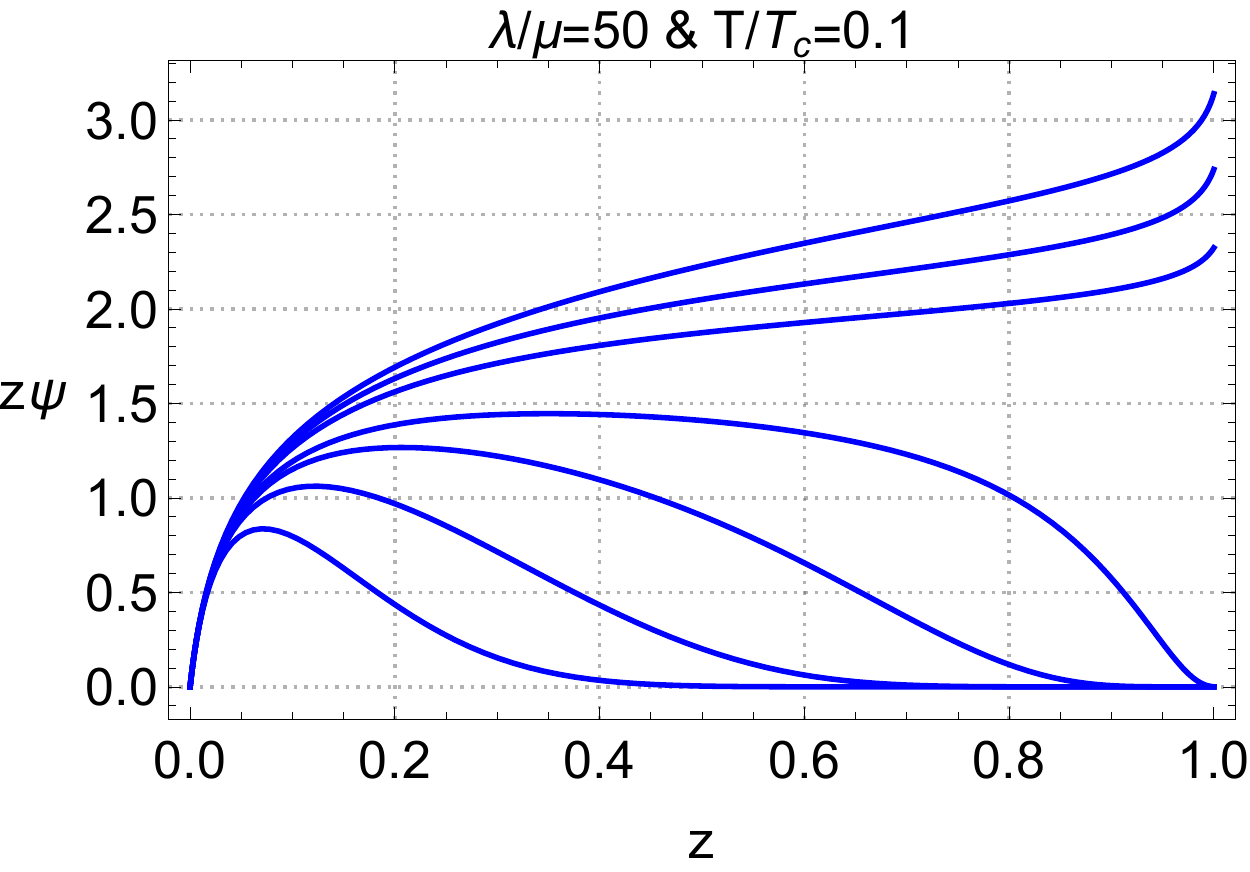}	
	\caption{The bulk profile of $|\Phi| = z \psi (z)$. From top to down the curves represent $k/\mu=2,4,6,10.1,12,15,20$.     $T/T_c=0.1$ and $\lambda/\mu=50$. }\label{fig3}
\end{figure}

Notice that both $T_c$ and $\sigma_{DC}(T_c)$ approach their values of the AdS-RN as $k/\mu \rightarrow \infty$. Indeed, as shown in the following section, the superfluid density also approaches the value of  the AdS-RN as $k/\mu \rightarrow \infty$.  This universal feature can be understood in two ways. First,  For $k/\mu \gg 1$,  $\Psi = z \psi e^{ikx}$ oscillates so fast that the lattice effect is averaged out and translational symmetry is effectively restored.  Second, the bulk profile of $|\Psi(z)| = z\psi$ becomes suppressed for $k/\mu \gg 1$ as shown in Figure \ref{fig3}, where,   for example, $|\Psi(z)|$ at $T/T_c=0.1$ with $\lambda/\mu=50$ is plotted for different $k/\mu$. 
For large $k/\mu$, $|\Psi(z)|$ are almost zero near horizon ($z=1$) so infrared physics will not be affected by $\Psi$.  

Figure \ref{fig3} also shows  that there is a qualitative change of $|\Psi(z)|$ at the critical value of $k_c/\mu \approx 10.1$. That is $|\Psi(1)| = 0$ for $k/\mu > k_c/\mu$ and $|\Psi(z)| \ne 0$ for $k/\mu < k_c/\mu$. Interestingly,  this critical $k_c/\mu$ when $\Phi \ne 0$ coincides with the MIT point when $\Phi=0$ in Figure \ref{MIT}.

In both Figures \ref{fig1} and \ref{fig2}, the curves have the solid part and the dotted part.  The former has the $k/\mu$ values of insulator   and the latter has the $k/\mu$ values of metal, where $k/\mu$ is read off from Figure \ref{MIT}.  It shows that $k/\mu$ dependence of $T_c$ and $\sigma_{DC}(T_c)$ has some correlation with the MIT. 

%%%%%%%%%%%%%%%%%%%%%%%%%%%%%%%%%%%%%%%%%
%%%%%%%%%%%%%%%%%%%%%%%%%%%%%%%%%%%%%%%%%

\section{Superfluid density} \label{sec4}

In this section we compute the superfluid density $\rho_s$ in two ways based on the London equation~\cite{Hartnoll:2008kx}: 
\begin{equation}\label{lon}
J_i(\omega, \vec{p})=-\rho_sA_i (\omega, \vec{p}) \,,
\end{equation}
which is valid when $\omega$ and $\vec{p}$ are small compared to the scale at which the system loses its superconductivity.
We will consider two limits: 1) $\vec{p}=0$ and $\omega \rightarrow 0$, 2) $\omega=0$ and $\vec{p} \rightarrow 0$. The two cases can explain the infinite DC conductivity and the Meissner effect of superconductors respectively.

First, in the limit $\vec{p}=0$ and $\omega \rightarrow 0$, the time derivative of (\ref{lon}) gives
\begin{equation}\label{lon1}
J_i(\omega,0)=\frac{i\rho_s}{\omega}E_i(\omega,0) \equiv \sigma(\omega) E_i (\omega, 0) \,,
\end{equation}
where $\sigma(\omega)$ denotes complex optical conductivity.  Thus the superfluid density is identified with the coefficient of $1/\omega$ pole in the imaginary part of the complex electric conductivity 
\begin{equation} \label{KsDef1}
\mathrm{Im}[\sigma(\omega)]=\frac{\rho_s}{\omega} + \cdots\,, 
\end{equation}
which implies the infinite DC conductivity (the delta function in the real part of the conductivity)
\begin{equation} \label{KsDef2}
\mathrm{Re}[\sigma(\omega)] =  \frac{\pi}{2}\rho_s\delta(\omega) \,,
\end{equation}
by the Kramers-Kronig relation
\begin{equation}
 \mathrm{Im}[\sigma(\omega)] =   -\frac{2\omega}{\pi} \mathcal{P} \int_0^{\infty} \dd\tilde{\omega} \frac{{\mathrm{Re}}[\sigma(\tilde{\omega})]}{{\tilde{\omega}}^2 - {\omega}^2} \,.
\end{equation}

The appearance of the delta function in $\mathrm{Re}[\sigma(\omega)]$ at $\omega =0$ in the superconducting phase is understood as the spectral weight transferred from finite $\omega$ by the Ferrell-Glover-Tinkham (FGT) sum rule \cite{Erdmenger:2015qqa,Kim:2015dna}
\begin{equation} \label{FGTeq}
\int^{\infty}_{0^+} \dd \omega \, \mathrm{Re} [\sigma_n (\omega) - \sigma_s (\omega)] = \frac{\pi}{2} \rho_s  \,,
\end{equation}
where $\sigma_n$ and $\sigma_s$ denote the electric optical conductivity in the normal phase and superconducting phase respectively. Physically, it means that the charged degrees of freedom of the system are conserved.

Second, in the limit $\omega=0$ and $\vec{p} \rightarrow 0$, the curl of (\ref{lon}) gives $\nabla \times \vec{J} = -
\rho_s \vec{B}$.
With Maxwell's equation $\nabla \times \vec{B} = 4 \pi \vec{J}$, we have
%\begin{eqnarray}
%-\nabla^2B =  -4\pi\tilde{K_s}B,\\
%\nabla^2B&=&\frac{1}{\lambda^2}B.
%\end{eqnarray}
\begin{equation} \label{mei}
\begin{split}
-\nabla^2 \vec{B}&=\nabla \times (\nabla \times \vec{B}) \\ 
&=4\pi\nabla \times \vec{J}=-4\pi{\rho_s}\vec{B} \equiv -\frac{1}{\lambda^2} \vec{B} \,,
%\nabla^2B&=&\frac{1}{\lambda^2}B.
\end{split}
\end{equation}
 implying the Meissner effect. Here $\lambda^2$ is the magnetic penetration depth squared which is inversely proportional to the superfluid density.  

\subsection{Holographic methods}

Based on these two limits, the superfluid density can be obtained experimentally by measuring optical conductivity or magnetic penetration depth.  Corresponding to both cases there are holographic computational methods.  According to the AdS/CFT correspondence $A_i$ and $J_i$ in \eqref{lon} are identified with the leading term $a_i^{(0)}$ and the sub-leading term $a_i^{(1)}$ in the expansion of the bulk gauge field $a_i(z)$ near  boundary $z=0$:
\begin{eqnarray}
a_i(z, \omega, \vec{p})&=&a_i^{(0)}(\omega, \vec{p})+z a_i^{(1)}(\omega, \vec{p})+\cdots \,.
\end{eqnarray}
Thus
\begin{equation}
\rho_s= \left. -\frac{a_i^{(1)}(\omega, \vec{p})}{a_i^{(0)}(\omega, \vec{p})} \right|_{\{\omega,\vec{p}\} \rightarrow 0}
\end{equation}
We can compute this by choosing a different limit  1) $\vec{p}=0$ and $\omega \rightarrow 0$, 2) $\omega=0$ and $\vec{p} \rightarrow 0$ corresponding to the optical conductivity and the magnetic penetration depth respectively\footnote{Since the gauge field in the holographic model is external, currents do not source electromagnetic fields and Maxwell's equation can not be applied in \eqref{mei}, but we still have a London equation.}. 
However, there is a subtle issue in the order of limit. The two limits $\omega\rightarrow0$ and $\vec{p}\rightarrow 0$ may not commute. In the probe limit, it was shown that the two limits  commute~\cite{Herzog:2009xv}, but in the case of full back reaction as in our set-up, these two limits may not commute. Because of this potential subtlety we will introduce new notations for superfluid density: $K_s$  for the case 1) and $\tilde{K}_s$ for the case 2).

First, to calculate the superfluid density in the limit $\vec{p}=0$ and $\omega\rightarrow0$, we introduce a small fluctuation of the gauge field of the form \cite{Ling:2014laa, Andrade:2014xca} 
\begin{equation}
\delta A_x=e^{-i \omega t}a_x(z) \,,
\end{equation} 
which is coupled to the fluctuations of the metric and the scalar field $\Psi$:
\begin{equation}
\delta g_{tx}=e^{-i \omega t}h_{tx}(z),~~~\delta \Psi=ie^{-i \omega t}e^{ikx}z\chi(z) \,.
\end{equation}
The equations of motion for $a_x(z), h_{tx}(z)$ and $\chi(z)$ are shown in appendix \ref{app1}.
Near boundary the asymptotic behaviour of the fluctuations are as follows:
\begin{eqnarray}
a_x(z)&=&a_x^{(0)}+z a_x^{(1)}+\cdots,\\
\chi(z)&=&\chi^{(0)}+z \chi^{(1)}+\cdots,\\
h_{tx}(z)&=&\frac{h_{tx}^{(0)}}{z^2}+\cdots.
\end{eqnarray}
We want to read off the electric conductivity only with electric field turned on, i.e. $\chi^{(0)}=h_{tx}^{(0)}=0$ However, as explained in detail in \cite{Donos:2013eha} if we impose ingoing boundary conditions near horizon it turns out that the number of independent parameters becomes only two, one of which should be $a_x^{(0)}$. Thus we cannot set both $\chi^{(0)}$ and $h_{tx}^{(0)}$ to be zero. However, if we impose $\omega\chi^{(0)}-ik\lambda h_{tx}^{(0)}=0.$ we may turn off the other sources by using diffeomorphism~\cite{Donos:2013eha}. With this condition we get 
\begin{equation} \label{k1}
\rho_s  = \left.-\frac{a_x^{(1)}(\omega, 0)}{a_x^{(0)}(\omega, 0)}\right|_{\omega \rightarrow 0} \equiv K_s \,,
\end{equation}
{which is equivalent to \eqref{lon1} because $i\omega a_x^{(0)}(\omega, 0) = E_x$ and $a_x^{(1)}(\omega, 0) = J_x$ by the AdS/CFT correspondence. }
%
%\begin{equation}
%	\delta A_y=e^{-i \omega t}a_y(z),~~~\delta g_{ty}=e^{-i \omega t}h_{ty}(z).
%\end{equation}

%\begin{eqnarray}
%	0&=&a_y^{''}+\left[\frac{((1-z)U)^{'}}{(1-z)U}+\frac{1}{2}\left(\frac{V_1^{'}}{V_1}-\frac{V_2^{'}}{V_2}\right)\right]a_y^{'}+\frac{\omega^2-2q^2(1-z)U\phi^2}{(1-z)^2U^2}a_y\\ \nonumber
%	&&+\frac{z^2((1-z)a)^{'}}{(1-z)U}h_{ty}^{'}+\frac{z((1-z)a)^{'}(2V_2-zV_2^{'})}{(1-z)UV_2}h_{ty},\\
%	0&=&\frac{z^2((1-z)a)^{'}}{(1-z)U}h_{ty}^{'}+\frac{z((1-z)a)^{'}(2V_2-zV_2^{'})}{(1-z)UV_2}h_{ty}+\frac{z^2((1-z)a)^{'2}}{(1-z)U}a_y.
%\end{eqnarray}
%
%\begin{equation}
%	0=a_y^{''}+\left[\frac{((1-z)U)^{'}}{(1-z)U}+\frac{1}{2}\left(\frac{V_1^{'}}{V_1}-\frac{V_2^{'}}{V_2}\right)\right]a_y^{'}+\frac{\omega^2-2q^2(1-z)U\phi^2}{(1-z)^2U^2}a_y-\frac{z^2((1-z)a)^{'2}}{(1-z)U}a_y.
%\end{equation}
%
%%%%%%%%%%%%%%%%%%%%%%%%%%%%%%%%%%%%%%%%%
 \begin{figure*}[]
 	\centering
 	\subfigure[\ No momentum relaxation]
 	{\includegraphics[scale=0.6]{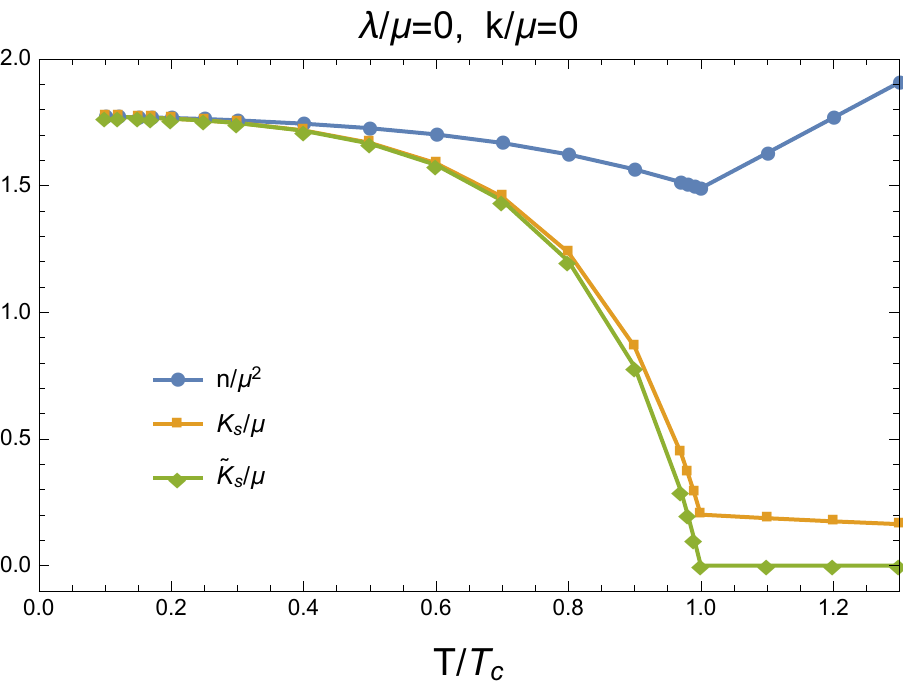} \label{fig4-1}}  \ \ \ \ \ \ \  \ \ \ \ \ \ \  
 	\subfigure[\ No momentum relaxation]
 	{\includegraphics[scale=0.6]{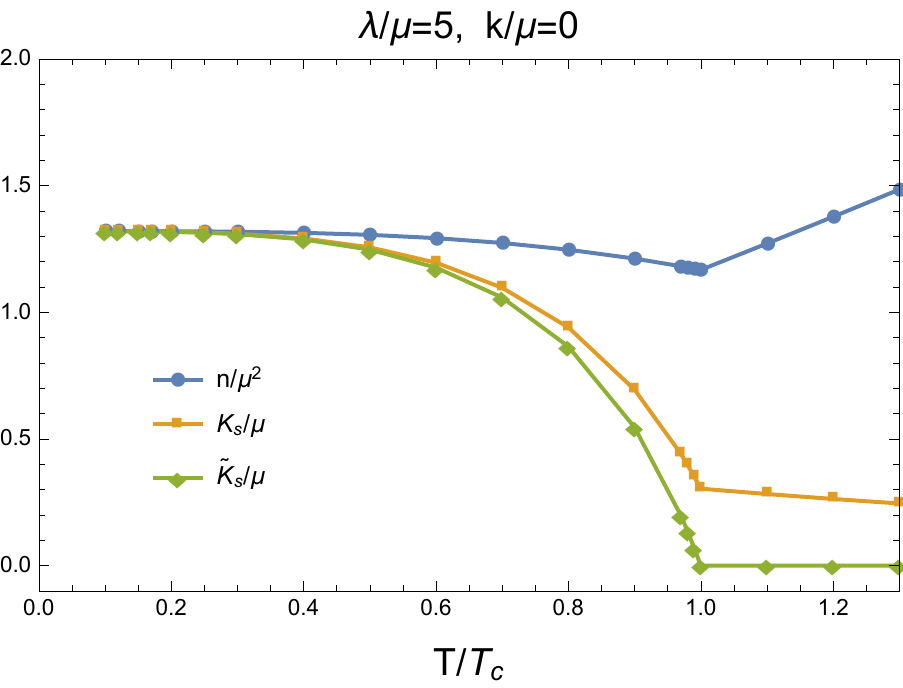} \label{fig4-2}}  \\
	 \subfigure[\ Large momentum relaxation]
 	{\includegraphics[scale=0.6]{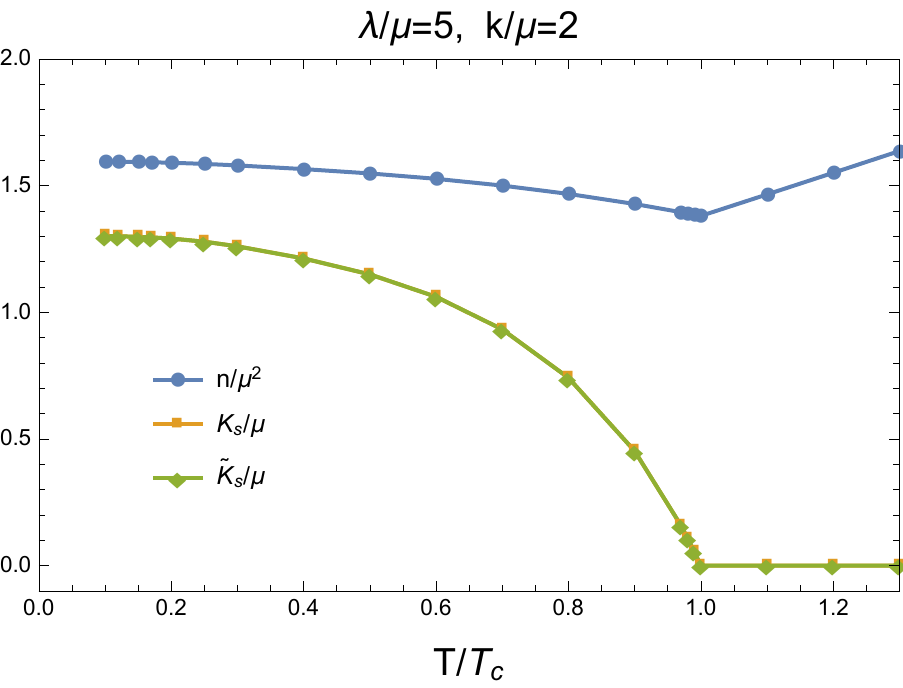} \label{fig4-3}} \ \ \ \  \ \ \ \ \ \ \ \ \ \  
 	\subfigure[\ Small momentum relaxation]
 	{\includegraphics[scale=0.6]{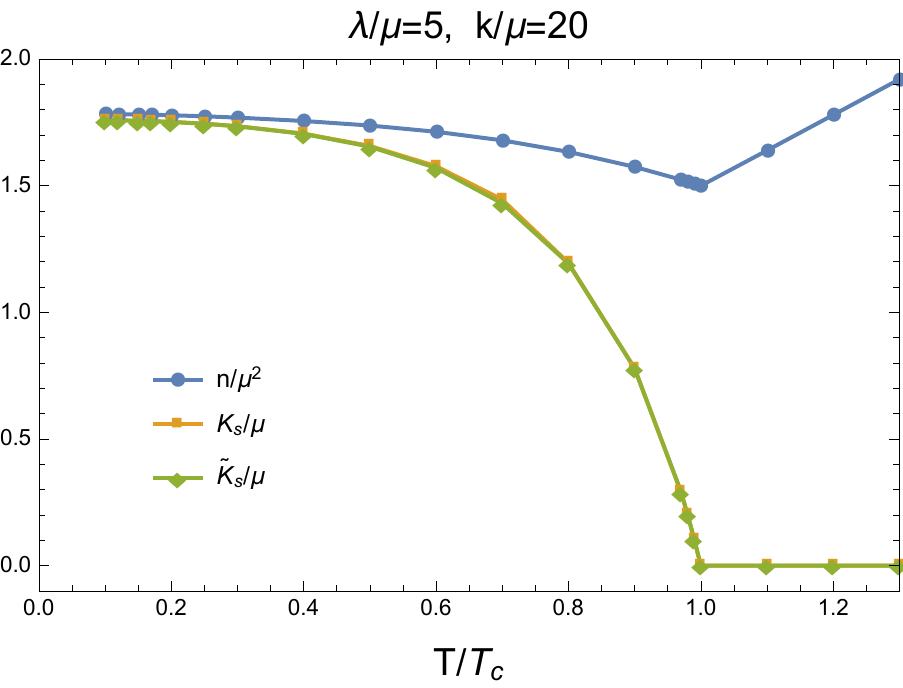} \label{fig4-4}}    
% 	\subfigure[]
% 	{\includegraphics[scale=0.8]{figs2} \label{fig4-3}}
% 	\subfigure[]
% 	{\includegraphics[scale=0.8]{figs3} \label{fig4-4}}  
%	 \subfigure[]
% 	{\includegraphics[scale=0.8]{figure6_c} \label{fig4-3}}
% 	\subfigure[]
% 	{\includegraphics[scale=0.8]{figure6_d} \label{fig4-4}}  
	\caption{The charge density $n$, superfluid density $K_s$, and $\tilde{K_s}$ vs $T/T_c$ }\label{Ks1}
 \end{figure*}
%%%%%%%%%%%%%%%%%%%%%%%%%%%%%%%%%%%%%%%%%
%
%%%%%%%%%%%%%%%%%%%%%%%%%%%%%%%%%%%%%%%%%
 %\begin{figure*}[]
% 	\centering
% 	\subfigure[\ no momentum relaxation]
% 	{\includegraphics[scale=0.6]{pfig1} \label{fig4-1}}  \ \ \ \ \ \ \ 
% 	\subfigure[\ no momentum relaxation]
% 	{\includegraphics[scale=0.6]{pfig2} \label{fig4-2}}  \\
%	 \subfigure[\ large momentum relaxation]
% 	{\includegraphics[scale=0.6]{pfig3} \label{fig4-3}} \ \ \ \  \ \ \
% 	\subfigure[\ small momentum relaxation]
% 	{\includegraphics[scale=0.6]{pfig4} \label{fig4-4}}    
% 	\subfigure[]
% 	{\includegraphics[scale=0.8]{figs2} \label{fig4-3}}
% 	\subfigure[]
% 	{\includegraphics[scale=0.8]{figs3} \label{fig4-4}}  
%	 \subfigure[]
% 	{\includegraphics[scale=0.8]{figure6_c} \label{fig4-3}}
% 	\subfigure[]
% 	{\includegraphics[scale=0.8]{figure6_d} \label{fig4-4}}  
%	\caption{The charge density $n$, superfluid density $K_s$, and $\tilde{K_s}$ vs $T/T_c$ }\label{Ks1}
 %\end{figure*}
%%%%%%%%%%%%%%%%%%%%%%%%%%%%%%%%%%%%%%%%%

Next we study the limit $\omega=0$ and $\vec{p}\rightarrow0$. In this case we introduce a fluctuation in $A_x$ that have momentum dependence of the form \cite{Maeda:2008ir}
\begin{equation}
\delta A_x=e^{i p y}a_x(z) \,.
\end{equation}
{Unlike~\cite{Maeda:2008ir}, we consider the back-reaction so $\delta A_x$ is coupled to the metric fluctuation:}
\begin{equation}
\delta g_{tx}=e^{i p y}h_{tx}(z).
\end{equation}
The equations of motions for these two fluctuations are written in appendix \ref{app1}. 
Near boundary
\begin{eqnarray}
a_x(z)&=&a_x^{(0)}+z a_x^{(1)}+\cdots \,,\\
h_{tx}(z)&=&\frac{h_{tx}^{(0)}}{z^2}+\cdots \,,
\end{eqnarray}
and setting $h_{tx}^{(0)}=0$ we have
\begin{equation} \label{k2}
\rho_s = \left.-\frac{a_x^{(1)}(0, p)}{a_x^{(0)}(0,p)}\right|_{p \rightarrow 0} \equiv \tilde{K}_{s} \,.
\end{equation}

\subsection{Numerical results}

Using \eqref{k1} and \eqref{k2} we have computed $K_s$ and $\tilde{K}_{s}$ as functions of $T/T_c$ for different sets of parameters $\lambda/\mu$ and $k/\mu$.  For example, in Figure \ref{Ks1}, we show our results for four cases: $(\lambda/\mu, k/\mu) = (0,0), (5,0), (5,2), (5,20)$. The orange curves are for $K_s/\mu$ and the green curves are for $\tilde{K}_s/\mu$. The blue curves represent the charge density $n/\mu^2\, $\footnote{The charge density is defined by a sub-leading term of $A_t$ in \eqref{sol}. i.e. $A_t = \mu - n z +\cdots$ near boundary.},  which is added for comparison.  

First, we display the cases with no momentum relaxation in Figure \ref{Ks1} (a) and (b): (a) is the case of AdS-RN geometry because $\lambda/\mu=0$ means $\Psi(z)=0$.  (b) is not AdS-RN, since there is a finite scalar field $\Psi(z)$ with a boundary value $\psi(0)/\mu =5$. However, the boundary theory is still translationally invariant because $k=0$.
Here we find that $K_s \ne \tilde{K}_s$ in general.  We expect the superfluid density vanishes $T > T_c$ so the superfluid density should be identified with $\tilde{K}_s$. The non-zero  $K_s$ for $T > T_c$ may be interpreted as a spurious effect 
by the infinite DC conductivity due to translational invariance. 
This is an interesting and useful observation, since $\tilde{K}_s$ gives a direct way to compute the superfluid density even in the case with translation invariance.

Next, let us turn to the case with momentum relaxation in Figure \ref{Ks1} (c) and (d). Here $K_s = \tilde{K}_s$ and they are zero for $T > T_c$, which means that the aforementioned spurious contribution to $K_s$  by translational invariance vanishes. 
%For the y-direction translational symmetry is preserved so $K_{sy} \ne \tilde{K}_{sy}$ and they are qualitatively the same as Figure \ref{Ks1}. 
Notice that the superfluid density $\tilde{K}_s$ in Figure \ref{Ks1} (d) is similar to $\tilde{K}_s$  in Figure \ref{Ks1} (a).
It is because in the limit $k \rightarrow \infty$ the translation invariance is effectively restored as explained at the end of section \ref{sec3}.   In this limit the value of $\lambda$ becomes irrelevant and the geometry approaches to the AdS-RN not the one for Figure \ref{Ks1} (b).

\begin{figure}[]
	\centering
	\includegraphics[scale=0.65]{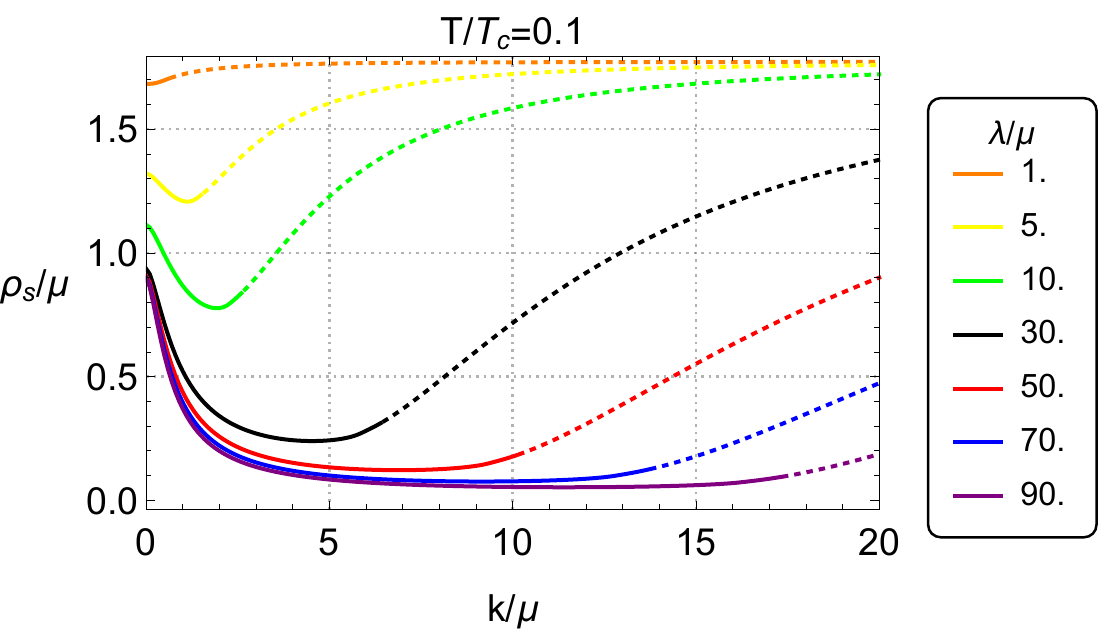}
	\caption{ Superfluid density $\rho_s/\mu (=K_s/\mu = \tilde{K}_s/\mu)$ at $T/T_c=0.1$  vs lattice wavenumber ($k/\mu$) at fixed lattice amplitude ($\lambda/\mu=1,5,10,30,50,70,90)$. The solid part and  the dotted part correspond to insulator and metal respectively in Figure\ref{MIT}.}\label{fig5-1}
\end{figure}

For our goal (Homes' law), we need to know $\rho_s$ at zero $T$.
$\rho_s = K_s = \tilde{K}_s$ near zero temperature for all cases, so we will use the notation $
\rho_s$ for superfluid density.
For example, $\rho_s$ at zero $T$ can be read from Figure \ref{Ks1} (b),(c),(d), for $\lambda/\mu=5$ and $k/\mu = 0,2,20$ respectively. 
Because of numerical instability of our numerical analysis we have obtained data up to $T/T_c=0.1$ and extrapolated them to $T=0$.
We have done this analysis for a range of $\lambda/\mu$ and $k/\mu$ and our results are shown in Figure \ref{fig5-1}.
For a fixed $\lambda/\mu$, $\rho_s/\mu$ at zero $T$ decreases when $k/\mu$ is small and increases when $k/\mu$ is large. As $k/\mu \rightarrow \infty$, it approaches to the AdS-RN value regardless of $\lambda$. In the curves, the solid part has the $k/\mu$ values of insulator and the dotted part has the $k/\mu$ values of metal in Figure \ref{MIT}. 
Similarly to $T_c$ (Figure \ref{fig1}) and $\sigma_{DC}(T_c)$ (Figure \ref{fig2}), the $k/\mu$ dependence of the superfluid density has some correlation with the MIT.

%%%%%%%%%%%%%%%%%%%%%%%%%%%%%%%%%%%%%%%%%
\begin{figure*}[]
	\centering
         \subfigure[\ Contour plot of $C=\rho_s/(\sigma_{DC}T_c)$ in $\lambda/\mu$-$k/\mu$ plane ]
	{\includegraphics[scale=0.5]{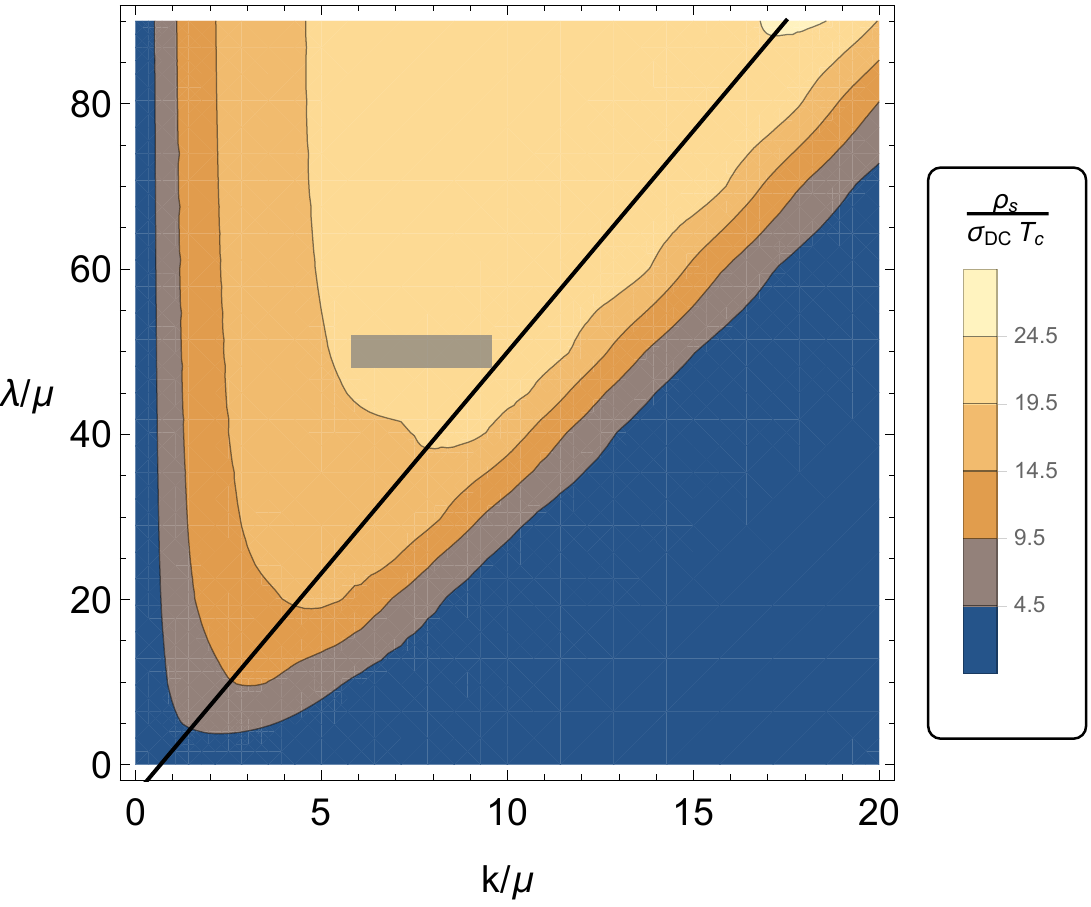}} \ \ \ \ \ \ \ \ \ \ 
	\subfigure[\ Cross-sections of (a) at $\lambda/\mu=1, 5, 10, 30,50,70,90$. The solid part and the dotted part correspond to insulator and metal respectively in Figure\ref{MIT}.]
	{\includegraphics[scale=0.65]{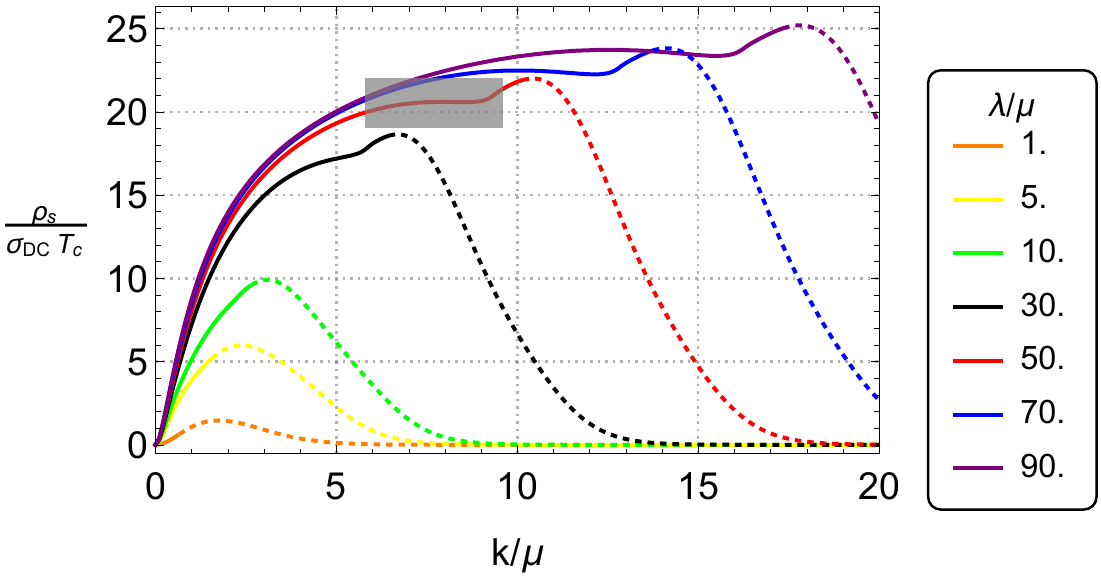}}
	\subfigure[\ Zoom-in of the grey window in (b) ]
	{\includegraphics[scale=0.6]{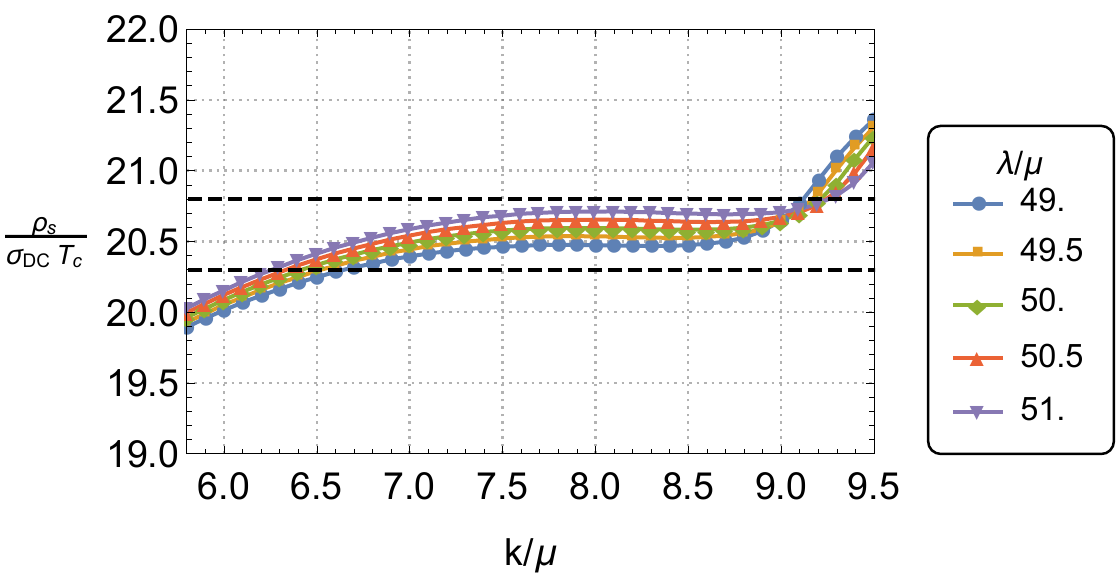} } \ \ \ \ \ \ \ \ \ \ \ \ \  \   
	\subfigure[\ $\rho_s/\mu$ vs $(\sigma_{DC}T_c/\mu)$: Log-Log plot for $\lambda/\mu \in (49,51)$ and $k/\mu \in (7,10)$.  ]
	{\includegraphics[scale=0.5]{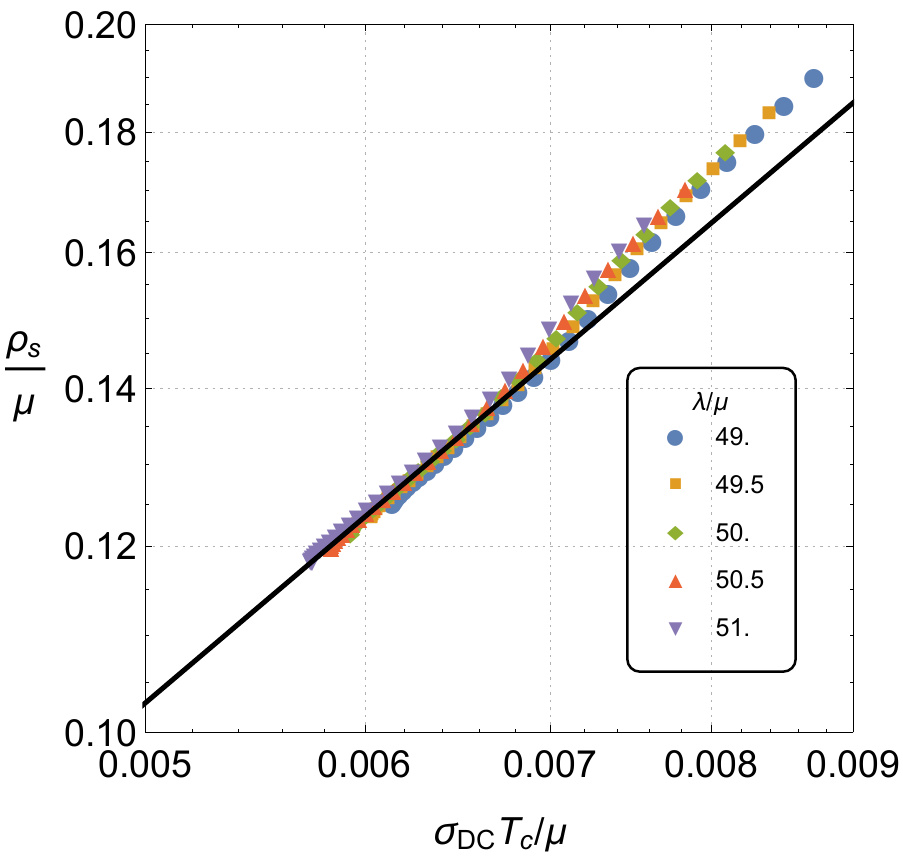} }  \ \ \ \ \ \ \ \ \ \  \    \ \ \ \ \ \ \ 
	\caption{Checking Homes' law: $C=\rho_s/(\sigma_{DC}T_c)$ as functions of $\lambda/\mu \in (0,90)$ and $k/\mu \in (0,20)$.  }\label{Fig:Homes1}
\end{figure*}

At zero temperature, without momentum relaxation $K_s/\mu=n/\mu^2$ (Figure \ref{Ks1}(a)(b)) while with momentum relaxation $K_s/\mu \ne n/\mu^2$ (Figure \ref{Ks1}(c)). This difference was also observed in other holographic superconductor models with momentum relaxation~\cite{Erdmenger:2012ik, Kim:2016hzi}, so it seems a general feature of holographic superconductors. Because the FGT sum rule \eqref{FGTeq} still holds even with $K_s/\mu \ne n/\mu^2$ we may conclude that some of the low frequency spectral weight is transferred to finite frequencies rather than the delta function at zero frequency. 
As another possibility to explain $K_s/\mu \ne n/\mu^2$ at zero $T$~\cite{Erdmenger:2012ik}, it was argued that the identification of superfluid density in \eqref{KsDef1} or \eqref{k1} may not be correct and it was proposed to cross check it via the magnetic penetration depth, which is \eqref{k2}.
We have cross checked it in our model and find two methods agree, $K_s = \tilde{K}_s$.

\section{Homes' law} \label{sec5}

Homes' law is given by
\begin{equation}
\rho_s(T=0)=C\sigma_{DC}(T_c)T_c \,,
\end{equation}
where $C$ is a universal material independent constant. In our holographic model, $k$ and $\lambda$ correspond to the properties of material so we want to check if $C$ is constant irrespective of $k$ and $\lambda$. 
Having computed $T_c$ (Figure \ref{fig1}), $\sigma_{DC}(T_c)$ (Figure \ref{fig2}), and $\rho_s$ (Figure \ref{fig5-1}) as functions of $k$ and $\lambda$ we are ready to check Homes' law in our holographic superconductor model.

First, to have an overall picture, we present a contour plot of $C=\rho_s/(\sigma_{DC}T_c)$  in $\lambda/\mu$-$k/\mu$ plane ($0 \le k/\mu \le 20$ and $0 \le \lambda/\mu \le 90$) in Figure \ref{Fig:Homes1}(a). The black diagonal line is the MIT line in Figure \ref{MIT}. 
In general, in the region close to the MIT, $C$ is larger and below the MIT, $C$ becomes small quickly. $C$ vanishes as $k/\mu \ll 1$ and $k/\mu \gg 1$ because $\sigma_{\mathrm{DC}} \gg 1$ due to the restoration of translational invariance. 

For $\lambda/\mu \gtrsim 40$, in a triangular region surrounded by a contour, $C$ does not change much compared to the other region. In that region, there is a possibility that Homes' law hold. To see it more clearly we make  Figure \ref{Fig:Homes1}(b), which is  the cross-sections of Figure \ref{Fig:Homes1}(a) for fixed $\lambda/\mu = 1,5,10,30,50,70,90$. Here we can see plateaus in some range of $k$ for every $\lambda /\mu\gtrsim 50$, which means Homes' law holds in that regime. 
 The regime are in the insulating phase near the MIT line, which was also observed in a holographic superconductor with  helical lattice~\cite{Erdmenger:2015qqa}.
In our case, Homes' law seems to hold for a wider range of $\lambda/\mu$ than the helical lattice case, even thought $C$ is a little bit different for a different $\lambda/\mu$\footnote{For large $\lambda$, it seems that $C$ is approaching to the universal value. However, we could not confirm it due to numerical instability for $\lambda/\mu > 90$.}.
In Figure \ref{Fig:Homes1}(c), we zoom in the grey window in Figure \ref{Fig:Homes1}(b) for $49 \le \lambda/\mu \le 51$. Figure \ref{Fig:Homes1} (d) is the Log-Log plot of $\rho_s/\mu$ vs $(\sigma_{DC}T_c/\mu)$  for $49 \le \lambda/\mu \le 51$ and $ 7 \le k/\mu \le 10$.
Figure \ref{Fig:Homes1} (c) and (d) are similar to Figure 15 in \cite{Erdmenger:2015qqa}.

The appearance of the plateaus for large $\lambda/\mu$ in Figure \ref{Fig:Homes1}(b) may be qualitatively understood from Figure \ref{fig1}, \ref{fig2} and \ref{fig5-1}, where all three quantities $\rho_s$, $\sigma_{\mathrm DC}$, and $T_c$ show the same qualitative behaviour. 
At fixed $\lambda/\mu$, as $k/\mu$ increases, they decrease at small $k/\mu$ and reach their minimum values and again increase at large $k/\mu$. 
As $\lambda/\mu$ grows, their minimum values are saturating and the plateaus start developing around the minimum.  
Bigger the $\lambda/\mu$,  longer the ranges of $k/\mu$ for plateaus.

However, if we look at closely, the plateau of every $\rho_s$, $\sigma_{\mathrm DC}$, ann $T_c$ is not strictly flat. They are 
slightly increasing or decreasing, but the combination of them, $C$, shows a better plateau behaviour. 
To check it explicitly
we have made a plot for $B \equiv \rho_s/ T_c$ without $\sigma_{\mathrm{DC}}$ and found that $B$  is not as flat as $C$ shown in Figure \ref{Fig:Homes1}(c). Physically, this means that the Uemura's law\footnote{Uemura's law is $\rho_{s}(T = 0) = B \, T_{c}$, where $B$ is another universal constant independent of materials. It holds only for underdoped cuprates~\cite{Homes:2005aa,Homes:2004wv}:} does not hold in our model.

In addition, there are also plateaus at fixed $k/\mu$ for some range of $\lambda$. It can be seen from the almost vertical part of contour lines for $k/\mu \le 5$ in Figure \ref{Fig:Homes1}(a).

\section{Conclusion and discussions} \label{sec6}

We investigated Homes' law by computing the critical temperature ($T_c$), the DC conductivity at the critical temperature ($\sigma_{\mathrm{DC}}(T_c)$), and the superfluid density ($\rho_s$) in a holographic superconductor with Q-lattice. 
In this set-up Homes' law means that $C=\rho_s/(T_c \sigma_{\mathrm{DC}}(T_c))$ is independent of the amplitude ($\lambda$) and/or wavenumber ($k$) of Q-lattice. We find that Homes' law holds for a range of $k/\mu$ at every fixed $\lambda/\mu \gtrsim 50$. As $\lambda/\mu$ grows, $C$ tends to approach to some universal value. Homes' law holds in insulating phase near the metal insulator transition (MIT), where momentum relaxation is strongest.  Roughly speaking, i) for a given $\lambda/\mu$, there is $k/\mu$ near the MIT (say, $k_c/\mu$) which gives the maximum value of $C$, ii) if $\lambda/\mu$ increases $C$ becomes constant for a range of $k/\mu$ around the $k_c/\mu$.

To compute the superfluid density, we employed two methods. One is related to the infinite DC conductivity and the other is related to the magnetic penetration depth. With finite momentum relaxation both give the same results, which serves as a good cross-check of our computation. However, without momentum relaxation only the latter correctly captures the superfluid density. The former gets spurious contribution from the infinite DC conductivity due to translational invariance.

At zero temperature, with momentum relaxation $K_s/\mu\ne n/\mu^2$  while without momentum relaxation $K_s/\mu =  n/\mu^2$. It was observed in other holographic models. Because the FGT sum rule \eqref{FGTeq} still holds  it seems that some of the low frequency spectral weight is transferred to finite frequencies rather than the delta function at zero frequency. 

In this paper, we considered the case with $q=6$ and $m_\Psi^2 = - 2$ in detail. We have also checked Homes' law for a different $q$ ($q=2$) and obtained qualitatively the same result. If $m_\Psi^2$ increases, it is possible that the MIT does not occur and consequently Homes' law does not hold.  For example,  if $m_\Psi^2 = 0$ our model becomes similar to the massless scalar model and it was shown that there is no MIT and no Homes' law in that model if $\Psi$ does not have $z$ dependence~\cite{Kim:2016hzi}.

Homes' law in our model comes from the MIT and strong momentum relaxation. 
The MIT seems to be less relevant phenomenologically but strong momentum relaxation is encouraging since it is a property of incoherent metal regime where Planckian dissipation \eqref{ppp1} may occur~\cite{Hartnoll:2014lpa}.
However, it turns out that our model does not have a linear in $T$ resistivity in normal (strange metal) phase as shown in Figure \ref{MIT2}. Because the linear in $T$ resistivity is a universal property of the normal phase of high $T_c$ superconductors and may be related to the physics of Homes' law by the Planckian dissipation~\cite{Zaanen:2004aa}, it will be  important to study Homes' law in a holographic model having linear in $T$ resistivity such as \cite{Davison:2013txa} \cite{WIP}.

\appendix
\section{Equations of motion for superfluid density} \label{app1}
We present the equations of motion for superfluid density used in section \ref{sec4}. The first one is for
the case $\vec{p}=0$ and $\omega\rightarrow0$ and the second is for $\omega=0$ and $\vec{p}\rightarrow0$.

\begin{widetext}
\noindent 1. $\vec{p}=0$ and $\omega\rightarrow0$
\begin{eqnarray}
	0&=&a_x^{''}+\left[\frac{((1-z)U)^{'}}{(1-z)U}+\frac{1}{2}\left(\frac{V_2^{'}}{V_2}-\frac{V_1^{'}}{V_1}\right)\right]a_x^{'}+\left(\frac{\omega^2}{(1-z)^2U^2}-\frac{z^2((1-z)a)^{'}}{(1-z)U}\right)a_x\nonumber \\
	&&-\frac{2q^2\phi^2}{(1-z)U}a_x+\frac{2ikz^2((1-z)a)^{'}(\psi^{'}\chi-\psi\chi^{'})}{\omega}, \nonumber \\	
       0&=&h_{tx}^{'}+((1-z)a)^{'}a_x+\left(\frac{2}{z}-\frac{V_1^{'}}{V_1}\right)h_{tx}-\frac{2q^2\phi^2}{(1-z)U}a_x-\frac{2ik(1-z)U(\psi^{'}\chi-\psi\chi^{'})}{\omega}\,,  \\
	0&=&\chi^{''}+\left[\frac{((1-z)U)^{'}}{(1-z)U}+\frac{1}{2}\left(\frac{V_2^{'}}{V_2}+\frac{V_1^{'}}{V_1}\right)\right]\chi^{'}+\left(\frac{\omega^2}{(1-z)^2U^2}-\frac{k^2}{(1-z)UV_1}\right)\chi\nonumber \\
	&&+\frac{1}{z}\left[\frac{((1-z)U)^{'}}{(1-z)U}+\frac{1}{2}\left(\frac{V_2^{'}}{V_2}+\frac{V_1^{'}}{V_1}\right)\right]\chi+\frac{2-2(1-z)U}{(1-z)z^2U}\chi-\frac{ik\omega z^2\psi}{(1-z)^2U^2V_1}h_{tx}, \nonumber
\end{eqnarray}
2. $\omega=0$ and $\vec{p}\rightarrow0$
\begin{equation}
\begin{split}
0&=a_x^{''}+\left[\frac{((1-z)U)^{'}}{(1-z)U}+\frac{1}{2}\left(\frac{V_2^{'}}{V_2}-\frac{V_1^{'}}{V_1}\right)\right]a_x^{'}-\frac{p^2+2q^2V_2\phi^2}{(1-z)UV_2}a_x\\
&+\frac{z^2((1-z)a)^{'}}{(1-z)U}h_{tx}^{'}+\frac{z((1-z)a)^{'}(2V_1-zV_1^{'})}{(1-z)UV_1}h_{tx},\\
0&=h_{tx}^{''}+\frac{1}{2}\left(\frac{4}{z}-\frac{V_1^{'}}{V_1}+\frac{V_2^{'}}{V_2}\right)h_{tx}^{'}+((1-z)a)^{'}a_x^{'}+\frac{2q^2a\phi^2}{U}a_x\\
&+\left[\frac{z^2((1-z)a)^{'2}}{2(1-z)U}+\frac{z^4\left(\frac{(1-z)U}{z^2}\right)^{'}\left(\frac{V_1}{z^2}\right)^{'}}{(1-z)UV_1}-\frac{p^2z^2+2V_2(3+z^2(\psi^2+\phi^2))}{(1-z)z^2UV_2}\right]h_{tx}.
\end{split}
\end{equation}
\end{widetext}

\acknowledgments

We would like to thank Tomas Andrade for collaborations at an early stage of this project.  We also would like to thank  Johanna Erdmenger, Sean Hartnoll, Elias Kiritsis,  Yi Ling, Andy O'Bannon, and Koenraad Schalm for valuable discussions and correspondence. The work was supported by Basic Science Research Program through the National Research Foundation of Korea(NRF) funded by the Ministry of Science, ICT $\&$ Future Planning(NRF- 2014R1A1A1003220) and the GIST Research Institute(GRI) in 2016.

%\bibliography{/Users/FortOe/Dropbox/Research/Template/KyKimRefs}
%\bibliography{/Users/apple/Dropbox/Research/Template/KyKimRefs}
%\bibliography{KyKimRefs}
%\bibliographystyle{JHEP}

\providecommand{\href}[2]{#2}\begingroup\raggedright\endgroup

\end{document}